\documentclass[12pt,english,aps,article,showpacs,floats]{revtex4}
\usepackage[latin1]{inputenc}
\usepackage{graphicx}
\usepackage{amssymb}

\makeatletter

\usepackage{amssymb,amsmath,graphicx,epsfig}

\newcommand{\bee}{\begin{equation}}
\newcommand{\ee}{\end{equation}}
\newcommand{\beea}{\begin{eqnarray}}
\newcommand{\eea}{\end{eqnarray}}

\usepackage{babel}

\usepackage{babel}
\makeatother
\begin{document}

\title{The Landscape of String Theory\\ and \\ The Wave Function
of the Universe}

\author{R. Brustein\protect$^{(1)}$ and S. P. de Alwis\protect$^{(2)}$}
\affiliation{(1) Department of Physics, Ben-Gurion University,
Beer-Sheva 84105,
Israel \\
 (2) Perimeter Institute, 31 Caroline Street N., Waterloo, ON N2L
2Y5, Canada}
\affiliation{Department of Physics, University of Colorado, Box 390,
Boulder,
CO 80309.\\
 \\
 \texttt{e-mail: ramyb@bgu.ac.il,  dealwis@pizero.colorado.edu} }

\begin{abstract}

We explore the possibility that quantum cosmology considerations could provide a
selection principle in the landscape of string vacua. We propose that the universe
emerged from the string era in a thermally excited state and determine, within a
mini-superspace model, the probability of tunneling to different points on the
landscape. We find that the potential energy of the tunneling end point from which
the universe emerges and begins its classical evolution is determined by the
primordial temperature. By taking into account some generic properties of the
moduli potential we then argue that the tunneling to the tail of the moduli
potentials is disfavored, that the most likely emergence point is near an extremum,
and that this extremum is not likely to be in the outer region of moduli space
where the compact volume is very large and the string coupling very weak. As a
concrete example we discuss the application of our arguments to  the KKLT model of
moduli stabilization.
\end{abstract}

\pacs{11.25. -w, 98.80.-k.}


\maketitle

\section{Introduction}

\label{S:intro}

The existence of a multiverse of solutions to string theory makes it eminently
desirable to find a dynamical principle that would select a universe, or a subset
of the multiverse that has properties similar to ours. Such a principle might be
provided in the framework of quantum cosmology
\cite{Vilenkin:1983xq,Hartle:1983ai,Linde:1983cm}. The wave function of the
universe yields a probability distribution for the dynamical parameters of the
multiverse. Depending on the boundary condition chosen for the wave function, the
probability distribution is sharply peaked at a zero value of the cosmological
constant (CC) \cite{Hawking:1984hk} or is peaked at a large value of the CC
\cite{Linde:1983cm,Vilenkin:1984wp} giving rise to an inflationary universe. A
related and complementary approach is based on the statistics of solutions
\cite{Bousso:2000xa,douglas1,douglas2,douglas3}, one of whose outcomes is that the
number density of solutions is uniform as a function of the value of the CC.

In this paper we propose that the universe emerged from the string era in a
thermally excited state above the Hartle-Hawking  vacuum. We show that imposing
this boundary condition on the wave function of the universe leads, within a
mini-superspace model, to interesting restrictions on the allowed dynamics. High
temperature effects were first introduced in a related context, to the best of our
knowledge, by Vilenkin \cite{Vilenkin:1998rp} who suggested including them in the
context of the so-called tunneling boundary condition, but did not work out the
consequences. We will show in detail the relevance of these effects both to the
Hartle-Hawking (HH) wave function and to the tunneling wave function.

Our proposal is motivated by an interesting observation that was made recently by
Sarangi and Tye \cite{Sarangi:2005cs} (see also \cite{Firouzjahi:2004mx} and
\cite{Huang:2005wq}). They considered the tunneling amplitude as a function of the
CC $\Lambda$, and found that if rather than evaluating it from the HH wave function
\cite{Hartle:1983ai} that describes tunneling from nothing,
\begin{equation}
\Psi\simeq e^{3\pi/G\Lambda},
 \label{Hwave}
\end{equation}
a modified wave function is used,  then  the tunneling amplitude
\begin{equation}
\Psi\simeq e^{3\pi/G\Lambda-(a/2)(3\pi/G\Lambda)^{2}},
 \label{STwave}
\end{equation}
has a critical point (a maximum for the HH choice). The additional parameter $a$ is
a constant that depends on the string mass scale $M_{s}$, the Planck scale
$M_{p}=G^{-1/2}$ and the number of fluctuating degrees of freedom $n_{dof}$. The
critical CC that Sarangi and Tye found is
$\Lambda_{c}=4n_{dof}\left(\frac{M_{s}^{2}}{M_{p}^{2}}\right)M_{s}^{2}$.

The original argument by Hawking was that since the wave function (\ref{Hwave}) is
peaked at $\Lambda\rightarrow 0^{+}$ it explains why the observed value of the CC
today is zero. Of course at that time there was just an observational upper bound
on the CC, and it was assumed that the upper bound implies that the true value is
zero. The problem with this argument is that it predicts a large but empty
universe. Sarangi and Tye argue that their modified wave function (\ref{STwave})
predicts inflation and hence makes the choice of the HH boundary condition more
interesting.

We use a mini-superspace model for our discussion. In this model only the
time-dependent scale factor of the universe, and some time-dependent but
homogeneous fields are considered and all other degrees of freedom are ignored. We
regard the mini-superspace calculation as a toy model which by itself is a well
defined quantum mechanical problem that can be solved self-consistently and
hopefully incorporates some features of the complete quantum gravity calculation.
At the current stage of development an attempt to do a more complete quantum
gravity analysis seems to lead to the notorious problems of Euclidean quantum
gravity. Thus our calculation should be regarded as an indication of the true
solution of the full dynamical problem. For self-consistency we apply the quantum
tunneling picture in four dimensional mini-superspace only for large scale factors
and energies and temperatures that are small compared to the string scale and the
Kaluza-Klein scale. The existence of the primordial temperature allows us to do
this  at the price of not being able to say anything about the quantum origin at
zero scale factor. We regard the inclusion of primordial radiation as a
parametrization of our ignorance of the physics at the quantum origin.

We start our discussion in the simpler case for which the moduli, the dilaton $S$,
and the volume modulus $T$ are fixed by stringy effects. In this case the thermal
boundary conditions lead to results that are similar to those of Sarangi and Tye.
Then we take into account the dynamics of the moduli by assuming they are
stabilized as in the recent works on moduli stabilization by fluxes and
non-perturbative effects \cite{Giddings:2001yu,Kachru:2003aw}. Moduli stabilization
by fluxes was considered previously in
\cite{Becker:1996gj,Gukov:1999ya,Dasgupta:1999ss,Taylor:1999ii,Greene:2000gh}. We
show that in order to have a finite potential barrier and therefore a large
tunneling amplitude, the end point of the tunneling should be to a region of the
moduli potential that supports accelerated expansion. Taking into account
additional features of the moduli potentials we further show that the tunneling
takes place to a maximum or a saddle point that is within a bounded region in
moduli space. In particular, we find that the tunneling to the tail of the moduli
potentials, where they runaway to decompactification for example, is disfavored.
The reason for this is that these regions do not lead to sustained accelerated
expansion.

The organization of the paper is as follows. After a brief review of the wave
function calculus in sect.~\ref{S:wavefunction} we discuss in sect.~\ref{S:thermal}
thermal effects and their significance to the tunneling amplitude. In
sect.~\ref{S:tunnprob} we discuss the dependence of the tunneling amplitude on the
parameters and conclude that we must add the dependence on moduli. We follow our
own conclusion and  in sect.~\ref{S:WDWscalar} and sect.~\ref{S:moduli} where we
extend the analysis to a case with dynamical moduli. In sec.~\ref{S:KKLT} we
discuss in detail the application of our results to the model of Kachru, Kallosh,
Linde and Trivedi (KKLT). Section \ref{S:conclusions} contains our conclusions, and
in the appendix we compare our work to that of Sarangi and Tye.

\section{Review of the wave function calculus}

\label{S:wavefunction}

Let us first briefly review the calculation of the wave function of the universe in
the mini-superspace context. \footnote{For and alternative to the standard
arguments see \cite{Daughton:1998aa}.} Consider a theory with a dynamical CC
$\Lambda$ whose action is
\begin{equation}
S=\frac{1}{16\pi G}\int d^{4}x\sqrt{g}(R-2\Lambda).
 \label{action}
\end{equation}
Here all other dynamical fields are ignored and it is assumed that their effects
are incorporated into a single parameter $\Lambda$ that is essentially an
integration constant. The problem is further simplified by considering only
homogeneous and isotropic metrics that describe a closed universe. For such metrics
the line element is
\begin{equation}
ds^{2}=\sigma^{2}\left(-dt^{2}+a^{2}(t)d\Omega_{3}^{2}\right),
 \label{metric1}
\end{equation}
$\sigma^{2}=2G/3\pi$. For this class of models the action simplifies
considerably,
\begin{equation}
S=\frac{1}{2}\int dt(-a\dot{a}^{2}+a-\lambda a^{3}).
 \label{action2}
\end{equation}
All quantities in the action (\ref{action2}) are dimensionless, in
particular, the
dimensionless CC $\lambda=\sigma^{2}\Lambda/3=2G\Lambda/9\pi$. The
canonical
momentum conjugate to $a$ is $\Pi_{a}=-a\dot{a}$ and the classical
Hamiltonian
constraint is
\begin{equation}
H=-\frac{1}{2a}\left(\Pi_{a}^{2}+U(a)\right)\thickapprox 0.
 \label{hamiltonian}
\end{equation}
The last relation means that the Hamiltonian vanishes on the space of
classical
solutions. The potential $U(a)=a^{2}-\lambda a^{4}$ is positive for
$0<a<1/\sqrt{\lambda}$ and negative for $a>1/\sqrt{\lambda}$.

The quantum equation derived from the Hamiltonian constraint is the
Wheeler-deWitt (WDW) equation which replaces the Schroedinger equation
in quantum gravity. The momentum operator
$\hat{\Pi}_{a}=-i\partial/\partial a$
replaces the classical momentum,
\begin{equation}
\hat{H}\Psi[a]=\left(\hat{\Pi}_{a}^{2}+U(a)\right)\Psi(a)=0.
 \label{wdw1}
\end{equation}
 The WDW equation can be solved in the WKB approximation,
\begin{eqnarray}
\Psi_{\pm}[a] & \sim& e^{\pm iS_{c}}  =e^{\pm
i\int_{0}^{a}\Pi_{a}da}\nonumber \\
 & =& e^{\pm\frac{1}{3\lambda}\left[1-(1-\lambda a^{2})^{3/2}\right]}.
 \label{wave}
\end{eqnarray}
The last equality is valid only for $0<a<1/\sqrt{\lambda}$. In eq.~(\ref{wave}) we
have ignored the prefactors because they will not be important for the rest of the
discussion. The classical action $S_{c}$ is evaluated on a classical solution so
that  $\Pi_{a}=\partial S_{c}/\partial a=\sqrt{-U(a)}$. For $0<a<1/\sqrt{\lambda}$,
$\Pi_{a}$ is of course imaginary and the wave function is real. Hartle and Hawking
\cite{Hartle:1983ai} have imposed the no-boundary boundary condition which in this
context means that the positive sign should be taken in eq.~(\ref{wave}). Then the
under the barrier wave function near $a=1/\sqrt{\lambda}$ is given by the growing
exponential, and the probability of the universe emerging after tunneling through
the barrier is given by
\begin{equation}
P_{HH}[\lambda]\sim e^{2/3\lambda}.
 \label{HH}
\end{equation}

Hawking \cite{Hawking:1984hk} has argued that this distribution explains the
observed vanishing of the CC. At that time it was assumed that the observations
indicated that it was in fact zero. Now there is strong evidence that the expansion
of the universe is accelerated and therefore that the CC, or a similar form of dark
energy is small and positive. However, as many have pointed out, the probability
distribution (\ref{HH}) also predicts an empty universe - there would have been no
primordial inflationary stage.

The result (\ref{HH}) depends very sensitively on the nature of the state and in
particular on the boundary conditions. If one uses the Euclidean path integral to
define the state, the no-boundary proposal gives the ground state wave function.
The resulting distribution is rather surprising. It implies that the tunneling
amplitude increases exponentially when the barrier becomes larger. This is
certainly not what happens in laboratory tunneling experiments which show that the
tunneling amplitude decreases exponentially with the size of the barrier.

Linde \cite{Linde:1983cm} and Vilenkin \cite{Vilenkin:1983xq} have proposed
different boundary conditions that yield results that are similar to the
probability distributions in standard tunneling processes. Vilenkin has proposed
that only an outgoing wave should exist in the Lorentzian region for
$a>1/\sqrt{\lambda}$. This means that the universe has only an expanding component
whereas the HH wave function is a superposition of expanding and contracting
universes. The probability distribution that results from Linde's and Vilenkin's
proposals is
\begin{equation}
P_{L,V}[\lambda]\sim e^{-2/3\lambda}.
 \label{LV}
\end{equation}
This clearly favors a large CC. In fact it favors a situation for which the barrier
is as small as possible, and even no barrier at all! Since the semi-classical
theory is valid only up to some cutoff scale, which we will choose to be the string
scale $M_{s}$, this seems to imply that the universe is created in a state with
string scale CC. Linde has proposed recently \cite{Linde:2004nz} that a flat
compact universe may perhaps be more likely, since in this case there is no
barrier.

To discuss inflation it is necessary to include at least one inflaton field. This
is done for instance in \cite{Vilenkin:1987kf} where it is shown that the formulae
(\ref{HH},\ref{LV}) remain valid when one replaces the CC with the scalar potential
$V(\phi)$,  $\lambda \rightarrow 4\pi^{2}\sigma^{2}V(\phi)$, provided that it is
slowly varying, i.e. $|V^{-1}dV/d\phi|\ll 1.$  Obviously, these semi-classical
considerations are valid only for regions in field space where $|V(\phi)|\ll
M_{s}^{4}$. In \cite{Vilenkin:1987kf} various possible potentials are illustrated
with the corresponding probability distributions for the two cases
(\ref{HH},\ref{LV}). In all cases it is clear that the HH distribution will favor
tunneling into the lowest positive points in the potential,  while the LV
distribution will favor tunneling into the highest points of the potential allowed
by the cut off.

Vilenkin \cite{Vilenkin:1987kf} has argued that the above considerations lead to a
`prediction' of inflation from the LV wave function, in contrast to what is
obtained from the HH wave function. However the problem is that the conclusion is
cutoff dominated. Besides, the natural value of a cutoff would be close to the
Planck scale,  perhaps one or two orders of magnitude below, and this would be too
high to agree with observations. The predicted Hubble parameter during inflation is
$H\sim\sqrt{M_{s}^{4}/M_{p}^{2}}$. If the string scale is taken  to be an order of
magnitude below the Planck scale - which is the case for string compactifications
where there are no anomalously large extra dimensions, then $H\sim 10^{16}GeV$, in
conflict with the WMAP result that $H\lesssim10^{14}GeV$. To get agreement, the
cutoff scale would have to be at least two orders of magnitude below the Planck
scale.

We would like to make two remarks about the possible application of the wave
function calculus to string theory. First, in the string theoretic context
$V(\phi)$ is typically steep, so that the condition $|V^{-1}dV/d\phi|\ll 1$ is
obeyed only in the vicinity of the critical points of the potential. Thus, strictly
speaking the analysis can only compare the relative probability of tunneling to
different critical points. The LV wave function would predict tunneling to the
highest critical point whilst the HH wave function would favor the lowest positive
one. Second, recent work has shown that in string theory the acceptable solutions
that have potentials that can stabilize the moduli necessarily involve fluxes of RR
and NS-NS fields. The fluxes make the parameters of the potential, and in
particular the CC, discrete integration constants. In the string theory context one
should really consider a probability distribution that depends on all the moduli as
well as on the flux parameters. In contrast, the original Hawking argument was made
in a model without a scalar field. The CC was treated as a dynamical quantity
arising as an integration constant characterizing the flux of a four form field
strength.

\section{Thermal effects}

\label{S:thermal}

Our boundary condition proposal for the wave function of the universe can be stated
as follows: The universe emerges from the string era in a thermal state above the
Hartle-Hawking vacuum.  We propose that the decay of all the string excited states
has created a primordial thermal gas of radiation at a temperature that is somewhat
below the Hagedorn temperature. At this high temperature there would be in addition
to the massless states some population of a Boltzmann suppressed massive string
states, and perhaps also a gas of branes \cite{brandenberger}. These may behave as
pressureless matter, or have some other behavior. We will ignore such contributions
for simplicity, since we do not expect this to change the qualitative behavior that
we find. Within the context of string theory the effective field theory arguments
that we use make sense only at the end of the string era when the energy densities
in the universe are somewhat below the string scale.

Rather than evaluating the no-boundary thermal partition function, we evaluate the
Euclidean mini-superspace path integral with a modified effective potential that
includes the temperature corrections. The leading temperature correction is very
simple, a negative term proportional to the fourth power of the temperature is
added to the potential in the WDW equation. Equivalently, we can solve the WDW
equation with a modified potential, and even though one does not expect a coherent
wave function to describe a thermal state, the square of this WDW `wave function'
would have the interpretation of a density matrix that measures relative
probabilities.

Since the temperature scales as the inverse scale factor the energy density of the
radiation will be of the form $\rho_{RAD}=K/(\sigma^{4}a^{4})$ where $K$ is a
constant. The effective potential that goes into the WDW equation becomes
\begin{equation} U(a)-K=a^{2}-\lambda a^{4}-K.
\label{effpot}
\end{equation}
The potential barrier is now in the region limited by the roots of
$U(a)-K=0$,
$a_{-}<a<a_{+}$,
\begin{equation}
a_{\pm}^{2}=\frac{1}{2\lambda}\left(1\pm\sqrt{(1-4K\lambda}\right).
 \label{apm}
\end{equation}
Of course to have a barrier at all  the radiation term cannot be too large
$4K\lambda <1$. This condition needs to be  satisfied for  the semi-classical
theory to remain valid as argued below.

To keep the semi-classical effective field theory approach self-consistent the
highest radiation density in the region of interest should be less than string
scale. i.e. $\rho(a_{-})\sim K/\sigma^{4}a_{-}^{4}=CM_{s}^{4}$ with $C<1$. Using
the value $a_{-}^{2}\sim1/2\lambda$ from eq.(\ref{apm}) we get $K\sim
\frac{1}{4\lambda^{2}}  M_{s}^{4}\sigma^{4}$. Expressing $C$ in terms of the number
of degrees of freedom $n_{dof}$ in thermal equilibrium , $C=n_{dof}/c^{4}$ we have
our final expression for the radiation energy density,
\begin{equation}
\rho=\frac{n_{dof}}{c^{4}}\frac{1}{4\lambda^{2}a^{4}}M_{s}^{4}.
 \label{density}
\end{equation}
 The constant $c$ must satisfy $c^{4}>n_{dof}$ for the consistency
of these arguments. This energy density corresponds to an initial temperature at
$a=a_{-}$ of $T\sim M_{s}/c$ i.e. a temperature close to the Hagedorn temperature.
The same estimate can also be obtained by requiring that the initial entropy is
close to saturating the entropy bound as discussed in the appendix. The condition
for the existence of a barrier is thus equivalent in this context to the condition
that the cosmological constant be smaller than the string scale.

Including the contribution of the thermal radiation energy density
into the effective action (\ref{action2}), it becomes,
\begin{equation}
S=\frac{1}{2}\int dt\left(-a\dot{a}^{2}+a-\lambda
a^{3}-\frac{\nu}{a\lambda^{2}}\right),
\end{equation}
where
\begin{equation}
\nu=n_{dof}\frac{1}{9\pi^{2}c^{4}}\frac{M_{s}^{4}}{M_{p}^{4}}.
 \label{nu}
\end{equation}
 The corresponding classical Hamiltonian that replaces the one in
eq.(\ref{hamiltonian}) is
\begin{equation}
H=-\frac{1}{2a}(\Pi_{a}^{2}+U(a)-\frac{\nu}{\lambda^{2}})\thickapprox0.
 \label{hamiltonian2}
\end{equation}
The boundaries of the barrier can be rewritten as
\begin{equation}
a_{\pm}^{2}=\frac{1}{2\lambda}\left(1\pm\sqrt{1-\frac{4\nu}{\lambda}}\right).
 \label{aplusminus}
\end{equation}
It is important that for the range $\lambda\ll1$, $0<4\nu/\lambda<1$ the scale
factor is large under the barrier $a_{\pm}\gg 1$, so the semi-classical
mini-superspace calculation is self-consistent. Also since one expects
$n_{dof}\sim10^{2}-10^{3}$, the above restriction on $c$ means that it is about 10
but need not be much bigger. In fact it is reasonable to expect that the initial
temperature at the beginning of the classical evolution $T\sim M_{s}/c$ is close to
but not quite at the string scale. The initial volume of the universe as it emerges
from under the potential barrier into the classically allowed region is
$\sigma^{3}a_{+}^{3}\sim\sigma^{3}/\lambda^{3/2}$. The horizon volume on the other
hand is approximately $H^{-3}\simeq\sigma^{3}/\lambda^{3/2}$. Thus the universe
starts its classical evolution having a volume which is approximately one horizon
volume.

Now, in addition to the Lorentzian region to the right of $a_{+}$, there is also a
Lorentzian region to the left of $a_{-}$. This is depicted in Fig.~\ref{F1}. The
solution of the WDW equation in the WKB approximation is obtained by matching the
solutions on the boundaries of the three different regions,
\begin{eqnarray}
\Psi_{I} & =A_{+}e^{+i\Phi_{I}}+A_{-}e^{-i\Phi_{I}},\\
\Psi_{II} & =B_{+}e^{+\Phi_{II}}+B_{-}e^{-\Phi_{II}},\\
\Psi_{III} & =C_{+}e^{+i\Phi_{III}}+C_{-}e^{-i\Phi_{III}}.
\end{eqnarray}
The exponents are given by
\begin{eqnarray}
\Phi_{I}(a) & =\sqrt{\lambda} \int\limits_{0}^{a}
\sqrt{(a_{-}^{2}- a^{2})(a_{+}^{2}-a^{2})}, & \,0<a<a_{-} \\
\Phi_{II}(a) & =\sqrt{\lambda} \int\limits_{a_{-}}^{a}
\sqrt{(a^{2}-a_{-}^{2})(a_{+}^{2}-a^{2})}, & \, a_{-}<a<a_{+} \\
\Phi_{III}(a) & =\sqrt{\lambda} \int\limits_{a_{+}}^{a}
\sqrt{(a^{2}-a_{-}^{2})(a^{2}-a_{+}^{2})} & \, a_{+}<a.
\end{eqnarray}
In practice the matching has to be done with care since the boundaries of the three
regions are turning points where $E=U$, and therefore the WKB approximation breaks
down there.

The new Lorentzian region $a<a_{-}$ can clarify and resolve the debate about which
linear combinations to take inside the forbidden region, and which boundary
conditions to choose. If one puts boundary conditions in this region that
correspond to ``initial conditions" and not to ``final conditions" about the state
of the universe when it starts its classical evolution after tunneling then any
generic choice will effectively be equivalent to the HH choice as we now show. An
example of quantum cosmology with a Lorentzian region for small $a$ was considered
in the context of brane gravity in \cite{Davidson}.

A generic boundary condition in region $I$ that is not too far from a ``stationary
state" in the sense that it has comparable incident and reflected waves, will yield
some finite ratio of $B_{+}$ to $B_{-}$. Unless the coefficient of the increasing
exponential $B_{+}$ is tuned specifically to vanish or to be much smaller than
$B_{-}$ then the rising exponential will dominate the wave function inside the
barrier and at the beginning of the Lorentzian region. Hence in practice, the HH
boundary conditions can be taken, and will be a very good approximation to the
generic situation. Of course, if for some reason (for instance, in analogy with
$\alpha$-decay) one would like to impose, as advocated by Vilenkin, that there is
only an outgoing wave function in region III, then the coefficient $B_{+}$ would
need to be very small. In our setup, this would seem to be  a very special choice.

The tunneling amplitude is of the form $e^{\pm\Phi}$ with
\begin{eqnarray}
\Phi & =& \sqrt{\lambda}\int_{a_{-}}^{a_{+}}da
\sqrt{(a^{2}-a_{-}^{2})(a_{+}^{2}-a^{2})}\nonumber \\
& = & \frac{1}{\lambda} (1-\frac{4\nu}{\lambda})
\frac{1}{8}\int_{-\pi/2\sqrt{\lambda}}^{\pi/2\sqrt{\lambda}}
\frac{\sin^{2}(2\sqrt{\lambda}\tau}{a(\tau)}  d\tau
 \label{phase}
\end{eqnarray}
In order to get the second equality we have made the substitution
$a^{2}=a(\tau)^{2} \equiv(1+\sqrt{(1-\frac{4\nu}{\lambda}}
\cos(2\sqrt{\lambda}\tau))/(2\lambda)$. A simple estimate of the integral which is
quite sufficient for our purposes is obtained by assuming a triangular integrand
$\sqrt{U(a)-\nu/\lambda^2}$ whose height is the maximal height of the potential
barrier $\frac{1}{\sqrt{4\lambda}}\sqrt{1-\frac{4\nu}{\lambda}}$, and whose width
is $a_{+}-a_{-}=\frac{1}{\sqrt{\lambda}}\sqrt{1-\frac{4\nu}{\lambda}}$. The
resulting estimate for $\Phi$ is
\begin{equation}
\Phi=\frac{1}{4\lambda}\left(1-\frac{4\nu}{\lambda}\right).
 \label{simplestimate}
\end{equation}
A more sophisticated analysis shows that the integral can be expressed in terms of
complete elliptic integrals of the first and second kind. In the limit
$\nu/\lambda\rightarrow0$ the exact result reduces, as it should, to the previous
case (see eq.~(\ref{wave})) and its value is $8/3$, such that in this limit
$\Phi=\frac{1}{3\lambda}$. The exact expression is
\begin{equation}
\Phi=\frac{\sqrt{2}}{\lambda} (1-\frac{4\nu}{\lambda})
\frac{1}{\sqrt{1+\sqrt{1-\frac{4\nu}{\lambda}}}}
\frac{1}{2m}\left\{2\frac{m-1}{m}K[m] + \frac{2-m}{m}E[m]\right\},
 \label{elliptic}
\end{equation}
where
$m\equiv2\sqrt{1-\frac{4\nu}{\lambda}}/(1+\sqrt{1-\frac{4\nu}{\lambda}})$.

Let us now choose the HH $+$ sign for the logarithm of the wave function for the
reasons that were explained previously (we will discuss what happens when the other
sign is chosen later). In this case, the tunneling amplitude is maximized at the
maximum of $\Phi$. Using the simple estimate of eq.~(\ref{simplestimate}), we find
that when
\begin{equation}
\lambda=8\nu
\end{equation}
 $\Phi$ is maximized, and its value is
\begin{equation}
\Phi|_{max}=\frac{1}{64\nu}.
\end{equation}
Using a more accurate numerical evaluation of the exact expression
gives,
\begin{equation}
\lambda=5.26\,\nu,
\end{equation}
and
\begin{equation}
\Phi|_{max}=\frac{0.12}{\nu}.
 \label{phimax}
\end{equation}

The important point here is that the ratio of the radiation energy density to the
CC is some finite fixed numerical constant. This means that the initial radiation
energy density determines the value of the CC,
\begin{equation}
\Lambda\simeq\frac{5n_{dof}}{\pi c^{4}}
\frac{M_{s}^{2}}{M_{p}^{2}}M_{s}^{2}.
\end{equation}
It is clear from this formula that the result is sensitive to the initial
temperature and that the precise estimation of the value of $\nu/\lambda$ is not
particularly important. This is essentially the same result as the one obtained in
\cite{Sarangi:2005cs}, except that they have effectively put the constant $c=1$.
Our derivation shows that setting $c$ to unity is inconsistent, in that it would in
effect give at the barrier a radiation energy density that is greater than the
string energy density.

\section{The tunneling probability}

\label{S:tunnprob}

In order to discuss in a meaningful way the relative probabilities for tunneling
into different points in the landscape and the issue of whether or not inflation is
favored one really needs, in addition to the CC, to introduce the set of moduli
fields $\phi$ and their potential $V(\phi)$. If the moduli potential is not steep,
one could (following \cite{Vilenkin:1987kf}) take over the results of
sect.~\ref{S:wavefunction} with the substitution
$\lambda\rightarrow4\pi^{2}\sigma^{4}V(\phi)$. However, the string moduli
potentials are steep except in a limited domain around their extrema. Let us ignore
this for the moment and come back to this issue in section \ref{S:moduli}.

Let us consider the set of dynamical parameters in the potential that are
determined by the fluxes, gauge groups, etc., and denote them collectively by
$\beta$.  These vary from point to point in the landscape. Now the string to Planck
mass ratio depends on the moduli $\frac{M_{s}}{M_{p}}\left(\phi\right)$. Using the
above substitution, our previous maximization argument gives,
\begin{equation}
4\pi^{2}\sigma^{4}V_{max}(\phi;\beta)\simeq 5\nu =
\frac{5n_{dof}}{9\pi^{2}c^{4}}
\left(\frac{M_{s}}{M_{p}}\left(\phi \right)\right)^{4}.
 \label{vmax}
\end{equation}
It is clear from eq.~(\ref{vmax}) that the maximization of the tunneling amplitude
puts a constraint on the parameters $\beta$ and the values of the moduli fields to
which the universe tunnels.

Let us now observe what happens when  the LV sign for the under the barrier wave
function is chosen corresponding to the usual tunneling situation, with only
outgoing waves in the final Lorentzian region. Then within the class of models that
we consider: a closed universe with a positive CC, and radiation whose temperature
does not exceed the Hagedorn temperature, the maximum of the tunneling amplitude
becomes a minimum. Now the tunneling amplitude is maximized at the edge of
parameter space when $\lambda=4\nu$, exactly at the point that the barrier
disappears! In this case eq.~(\ref{vmax}) should be replaced by
\begin{equation}
4\pi^{2}\sigma^{4}V_{max}(\phi;\beta)=4\nu =
\frac{4n_{dof}}{9\pi^{2}c^{4}}
\left(\frac{M_{s}}{M_{p}}\left(\phi \right)\right)^{4}.
 \label{vmax2}
\end{equation}
Clearly, given that we are ignoring order one factors, the difference between the
two cases is not that significant. The thermal boundary condition switches the
physical consequences of  the two wave functions. In the absence of radiation the
LV wave function favors a larger CC whilst the HH wave function favors a zero CC.
With radiation the HH wave function favors a larger CC than the LV.

The maximization that we have performed was essentially with respect to $\lambda$
(or $V$) keeping $\nu$ fixed. In \cite{Sarangi:2005cs} probabilities for tunneling
for different values of $\nu$ are compared. However, if we strictly follow their
logic there appears to be a problem. From eq.~(\ref{phimax}) we have,
\begin{equation}
\Phi_{max}\simeq\frac{\sqrt{2}}{12\nu}
=\frac{\sqrt{2}}{12}\frac{9\pi^{2}c^{4}}{n_{dof}}\frac{M_{p}^{4}}{M_{s}^{4}}.
 \label{phimax2}
\end{equation}
The number of light degrees freedom does not change that much - being around
$10^{2}-10^{3}$ so the tunneling probability is essentially controlled by the ratio
of the Planck scale to the string scale. In
the heterotic string, for example, this is given by %
(See for example \cite{Polchinski:1998rr} chapter 18).
$\frac{M_{s}^{2}}{M_{p}^{2}}=\frac{\alpha_{YM}}{8}$ where
$\alpha_{YM}=g_{YM}^{2}/4\pi^{2}$ is the gauge field coupling strength at the
string scale. In type I theory on the other hand the ratio is given by (See for
example \cite{Brustein:2002mp}).
$\frac{M_{s}^{2}}{M_{p}^{2}}=g\frac{\alpha_{YM}}{4}$ where $g$
is the string coupling %
\footnote{Note that in our universe $\alpha_{YM}\simeq1/25$ so given that
$M_{p}\simeq10^{19}GeV$ we get a string scale of about $10^{17}GeV$ if the
heterotic string is the correct theory whilst a similar result is obtained if
$g\sim O(1)$ in the type I case too. This is the value we quoted earlier in the
discussion on inflation.%
}. Plugging these into eq.~(\ref{phimax2}) we get
\begin{eqnarray*}
\Phi_{max} & = & \frac{\sqrt{2}}{12}\frac{9\pi^{2}c^{4}}{n_{dof}}
\frac{8}{\alpha_{YM}} \,\,\,\,\,{\textrm{Heterotic}},\\
 & = & \frac{\sqrt{2}}{12} \frac{9\pi^{2}c^{4}} {n_{dof}}
 \frac{4}{g\alpha_{YM}} \,\,\,\,{\textrm{Type I}}.
\end{eqnarray*}
This seems to favor tunneling into very weakly coupled universes! If the value of
$M_{s}/M_{p}$ is not fixed then it is preferable to have it vanishingly small, and
then the radiation energy density also vanishes and consequently also the CC. The
final result in this case is very similar to the original HH result, the universe
tunnels to the smallest possible value of the CC. We believe that to come to any
reliable conclusion it is really necessary to explicitly consider the dependence on
the moduli scalar fields, which we do in the next sections.

\section{Action and Wheeler-de Witt equation for gravity and a scalar
field}

\label{S:WDWscalar}

As we have explained, to determine the tunneling probability it is necessary to
reexamine the wave function of the universe when in addition to the gravity sector,
we also have scalar fields. For simplicity we will consider just one field $\phi$
with the action
\begin{equation}
S_{\phi}=-\int d^{4}x\sqrt{g}
\left(\frac{1}{2}g^{\mu\nu}\partial_{\mu}\phi\partial_{\nu}\phi +
V(\phi)\right).
\end{equation}
For a closed universe the total mini-superspace action is
\begin{equation}
S=\frac{1}{2}\int dt\left\{ -a\dot{a}^{2}+
\sigma^{2}a^{3}\dot{\phi}^{2}+ a
-4\pi^{2}\sigma^{4}a^{3}V(\phi)\right\}.
\end{equation}
Here we have absorbed the CC into the scalar potential $V$. The
conjugate momenta
are
\begin{eqnarray}
\Pi_{a} & =-a\dot{a}\nonumber \\
\Pi_{\phi} & =a^{3}\sigma^{2}\dot{\phi},
\end{eqnarray}
and the Hamiltonian is
\begin{equation}
H=-\frac{1}{2a}\Pi_{a}^{2}+\frac{1}{2a^{3}\sigma^{2}}
\Pi_{\phi}^{2}-\frac{1}{2a}\left(a^{2}- 4\pi^{2}a^{4}\sigma^{4}V(\phi)
\right).
\end{equation}
The term $\frac{1}{2a^{3}\sigma^{2}}\Pi_{\phi}^{2}$ determines the kinetic energy
(KE) of the scalar field.

The WDW equation is $H\Psi(a,\phi)=0$, and can be obtain by the substitution
$\Pi_{a}\rightarrow-i\frac{\partial}{\partial a}$, and
$\Pi_{\phi}\rightarrow-i\frac{\partial}{\partial\phi}$. The result is then the
following,
\begin{equation}
\left[a^{2}\widehat{\Pi}_{a}^{2} - \frac{1}{\sigma^{2}}
\widehat{\Pi}_{\phi}^{2} +
a^{2}\left(a^{2} - 4\pi^{2}a^{4} \sigma^{4}V(\phi)\right)\right]
\Psi(a,\phi)=0.
 \label{wdweq}
\end{equation}
We have assumed a particular operator ordering but this is not particularly
important for our arguments.

\subsection{A slowly rolling scalar field}

\label{SS:slowroll}

In the limit that the scalar field is moving very slowly and where
$2\sigma^{4}V(\phi)$ can be treated as a constant, the WDW equation is separable
$\Psi(a,\phi)=\chi(\phi)\psi(a)$. Then,
\begin{eqnarray}
\frac{\partial^{2}\chi(\phi)}{\partial\phi^{2}} &
=&-\sigma^{2}E\chi(\phi)\\
a^{2}\frac{\partial^{2}\psi(a)}{\partial a^{2}}-
\left(a^{2}-a^{4}4\pi^{2}\sigma^{4}V(\phi)-\frac{E}{a^{2}}\right)\psi(a)
& =&0.
\end{eqnarray}
The significance of $E$ is clear: it represents the average KE of the scalar field.
The term $E/a^{2}$ in the WDW equation originates from a term $E/a^{6}$ in the
energy density, which is indeed a scalar field KE term.

Positive $E$ corresponds to positive KE, and leads to an oscillating scalar field
wave function while negative $E$ leads to an exponential scalar field wave function
whose interpretation is unclear. The solutions of the scalar field WDW are simply
linear combinations of  ``free particle" solutions: wave packets. For positive $E$,
denoting $k=\sqrt{E}$,
\begin{equation}
\chi(\phi)=\int\limits _{-\infty}^{\infty}dke^{\imath\sigma
k\phi}\tilde{\chi}(k).
\end{equation}
The choice $E=0$ leads to a constant wave function and a uniform probability for
all values of the field. If one chooses $E=0$, and the path integral is dominated
by a single classical configuration, then this classical configuration is a
constant scalar field, for consistency.

\subsection{A rolling scalar field}

\label{SS:fastroll}

If the potential is such that the field is moving significantly, we may use the
following method to solve the WDW equation. We find the classical solution for the
scale factor and scalar field, and express the scalar field as a function of the
scale factor. This is possible provided that the scale factor is a monotonic
function of time. We then use the parametric solution to find the kinetic and
potential energy of the scalar field as a function of the scale factor as shown
below. This results in a modified equation for the scale factor only.

The energy density of the scalar field
$\rho_{\phi}=\frac{1}{2}\dot{\phi}^{2}+V(\phi)$ and its pressure is
$p_{\phi}=\frac{1}{2}\dot{\phi}^{2}-V(\phi)$. The conservation equation
\begin{equation}
d\rho_{\phi}+3(\rho_{\phi}+p_{\phi})d\ln a=0,
\end{equation}
which is valid also for a closed universe, can be formally integrated if an
additional relation between $\rho$ and $p$ is supplied,
\begin{equation}
\rho_{\phi}(a)=\rho_{\phi}(a_{0}) e^{-3\int\limits _{a_{0}}^{a}
\left(1+\frac{p_{\phi}}{\rho_{\phi}}\right) d\ln a}.
\end{equation}

Let us consider as a simple instructive example the Euclidean scaling solution for
which the total energy density of the field behaves as a fixed power of the scale
factor: $\rho_{\phi}=\frac{1}{2}\dot{\phi}^{2}+V(\phi)\sim C/a^{m}$, corresponding
to the case that the equation of state (EOS) parameter
$w_{\phi}(a)=p_{\phi}(a)/\rho_{\phi}(a)$ is constant. The constant $C$ replaces the
CC for the general case in which $\rho_{\phi}$ is not constant, and the solution is
the same as the one with a fluid whose EOS parameter is $w=p/\rho=m/3-1$. In
general, the scalar field will not have a constant EOS parameter.

In the presence of space curvature it is  difficult to find potentials that yield a
constant EOS. For the Lorentzian spatially flat case it is well known that
exponential potentials lead to scaling solutions. A potential of the form
$V(\phi)=Ae^{\alpha\phi}$ leads to solutions that have $\dot{\phi}^{2}\sim
V(\phi)\sim1/t^{2}$ and $w=-1+2\alpha^{2}/3$. For $\alpha<1$ the potential is flat
enough and the expansion is accelerated. The maximal EOS parameter for a scalar
field with a positive potential is $w=1$ in the limit that the evolution is KE
dominated. This limit is reached when $\alpha$ approaches the critical value
$\sqrt{3}$, and $A$ vanishes. For a scaling solution to exist for larger values of
$\alpha$, the prefactor A needs to be negative, and in this case values of the EOS
parameter that are larger than unity are allowed. A positive potential that is
steeper than the critical steepness leads very quickly to a KE dominated evolution,
as discussed for example, in \cite{bdam}. This will be relevant to our discussion
in sect.~\ref{S:KKLT}

The Euclidean equations are in fact the same as the Lorentzian equations for the
spatially flat case, and therefore the Euclidean scaling solutions lead to the same
$\rho_{\phi}(a)$ as do the Lorentzian ones.

In the WDW equation (\ref{wdweq}), the terms that correspond to $\rho_{\phi}$ are
$-\frac{1}{2a^{2}\sigma^{2}}\Pi_{\phi}^{2}-2a^{4}\sigma^{4}V(\phi)$, so on a
scaling solution we obtain the following effective WDW equation for the scale
factor only,
\begin{equation}
\left[-\frac{\partial^{2}}{\partial
a^{2}}+a^{2}-Ca^{4-m}\right]\Psi(a)=0.
 \label{wdwgeneq}
\end{equation}
One can immediately realize that for the potential barrier to exist and be finite,
one needs that $m<2$, so that the scaling behavior corresponds to a power law
inflation with an equation EOS $p/\rho$ that is more negative than that of spatial
curvature $p/\rho<-1/3$. In general, on a solution $\rho_{\phi}$ can be replaced by
some function of $a$. The potential is then $U(a)=a^{2}-\rho_{\phi}(a)$. Clearly
$\rho_{\phi}$ needs to grow at large $a$ faster than $a^{2}$, and therefore needs
to have an EOS more negative than that of curvature.

Recall that the EOS of space curvature is $-1/3$ which is just the borderline
between decelerated and accelerated expansion. Thus,  our conclusion is that the
universe tunnels to a state of accelerated expansion. The scalar field energy
density needs to dominate over the space curvature for a while, in order to get the
potential down to zero, where the Lorentzian era begins. This conclusion will be
very important for the following discussion about the landscape.

\begin{figure}
\begin{center}
\includegraphics[width=10cm]{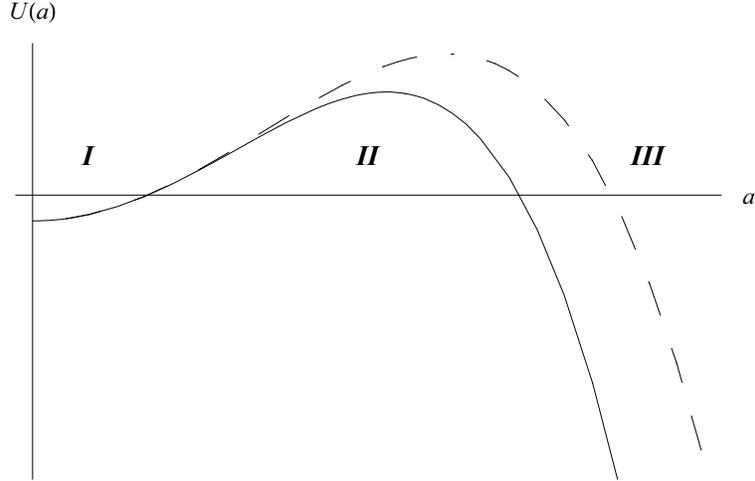}
\end{center}
\caption{The potential of the WDW equation for a closed universe with thermal
radiation. The solid line corresponds to the potential for a universe with a CC,
and the dashed line corresponds to a universe with dark energy with a softer EOS.
Regions I, III are classically allowed regions where the wave function is
oscillatory, while region II is a classically forbidden region where the wave
function is exponential.}
 \label{F1}
\end{figure}

An additional conclusion is that if the dominant power in $\rho_{\phi}$ has a less
negative EOS, the larger the range of $a$ that is required for the {}``dark
energy\char`\"{} to become dominant over the space curvature. This is depicted in
Fig.~\ref{F1}. In addition, for lower dominant power in $\rho_{\phi}$, the faster
the field $\phi$ is moving. As the evolution becomes less and less similar to that
in the presence of a CC, the scale factor is moving more when under the barrier and
the scalar field needs to move more. This requires that the potential of the scalar
field be flat over a larger region in field space so the accelerated expansion
could be supported. Another general comment is that one can compare the tunneling
amplitude for a CC to that in the case of  ``softer" EOS (less negative than $-1$).
Assuming that the EOS is a constant, leading to a power law dependence
$\rho_{\phi}\sim C/a^{(2-m)},m>2$, we can see that the tunneling amplitude that
comes form the positive exponential mode  (in the case that the magnitudes of the
potentials are similar) is much larger for the softer EOS, since the barrier is
much larger. This is depicted in Fig.~\ref{F1}.

It should be stressed here that the considerations of this section are independent
of the specific choice of the boundary conditions on the wave function. We have
argued that generic $initial$ conditions lead to the dominance of the rising mode
under the barrier. If however one imposes final conditions such  that only the
outgoing wave is allowed, then the usual tunneling picture emerges. Obviously, in
this case too the barrier needs to end and therefore, as argued above, the
tunneling needs to take place to an accelerating universe. By altering the topology
of the final state as in \cite{Linde:2004nz} (or even just the geometry by taking
flat or negatively curved universes) one could eliminate the barrier altogether and
in this case our arguments would be irrelevant.

\section{The tunneling probability with moduli}

\label{S:moduli}

The discussion and results of sections \ref{S:thermal} and \ref{S:tunnprob} about
the tunneling probability and its extremization did not take into account the fact
that the ratio of the string scale to the Planck scale $M_{s}/M_{p}$ is dynamical
and moduli dependent, and that the CC should be replaced by the moduli potential.
We reevaluate them in the light of the discussion of section \ref{S:WDWscalar}. We
make several assumptions about the form of the moduli potentials that allow us to
obtain more definite results.

In the case that the scalar field $\phi$ is a modulus, there are two possibilities.
First, that the true CC is large and then the potential is dominated by the CC.
This case is not very interesting because it does not lead to a universe that is
similar to ours. The more interesting possibility is that the true CC is
substantially less than the string scale. In the latter case we know from general
arguments about Peccei-Quinn symmetries and how they break, that the potential is a
sum of steep functions. Further, we know that at generic points that are not
extrema of the potential and that are in the outer region of moduli space, the
potential is dominated by a single steep function (see, for example, a related
discussion in \cite{bda1}). It follows that the only flat regions in the potential
of the moduli fields where $V'/V \ll 1$ occur in the vicinity of an extremum.

We argue that consequently, when the CC is small the universe cannot tunnel to a
generic point in the outer region of moduli space. We have seen that the tunneling
end point needs to be where the potential energy dominates the energy balance in
the universe, and in particular dominates the field's KE. However, we have just
argued that the potential for a generic point in the outer region of moduli space
is  steep.  On the other hand, we know that the cosmological solution of a scalar
field on a steep potential leads to the dominance of KE over the potential energy.
Our conclusion follows. Additionally, we have just argued that the only flat
regions in the potential of the moduli fields are in the vicinity of an extremum.
Our tentative conclusion is therefore that the universe tunnels to a region that is
not far from an extremum of the moduli potential. We proceed to examine this
conclusion in a more concrete setup.

A possible loophole in our argument could exist if some hitherto unknown Euclidean
transient solutions that support accelerated expansion for a short period of time
could be found. We expect such solutions, if they exists at all, to require some
special initial conditions. We cannot discuss their possible existence in the
general setup without adding more specific information on the moduli potentials. In
sect.~\ref{S:KKLT} we show that such solutions, even if they do exist, do not
modify our conclusions.

The moduli potential can be put in the form of an $N=1$ SUGRA potential,
\begin{equation}
V=e^{K}\left[D_{i}WD_{\bar{j}}\bar{W}K^{i\bar{j}}-3|W|^{2}\right]
 \label{potential}
\end{equation}
with
\begin{equation}
K=-\ln(2 S_R)-3\ln(2 T_R)+...
 \label{kaehler}
\end{equation}
Here $T_{R}$ is the real part of $T$ and $S_{R}$ is the real part of $S$ and the
ellipses represent the contribution of the other (Kaehler and complex structure)
moduli. The field $S$ is the (complex) dilaton axion field whose expectation value
determines the coupling and $T$ is the so-called volume modulus whose expectation
value determines the size of the internal manifold.  Together, they determine the
string to Planck mass ratio $M_{s}^{4}/M_{p}^{4}=S_{R}^{-1}T_{R}^{-3}$. Let us
denote the set of dimensionless moduli $(S,T,...)$ by $\phi^{i}$. The potential is
a function of these moduli as well as of the flux parameters and Casimirs of the
gauge groups which we collectively denoted by $\beta$ as in the discussion above
eq.~(\ref{vmax}). Thus we may express the potential as
\begin{equation}
V=V(\phi^{i};\beta).
\end{equation}
The WDW equation in the presence of the radiation term becomes,
 \begin{equation}
\left[-a^{2}\left(\frac{\partial}{\partial a}\right)^{2}+
K^{i\bar{j}}\frac{\partial^{2}}{\partial\phi^{i}\partial\phi^{\bar{j}}}+
a^{2}\left(a^{2}-a^{4}V(\phi;\beta)-K\right)\right]\Psi(a,\phi)=0.
 \label{WDWmoduli}
\end{equation}
Here we have substituted $\lambda\rightarrow V$ as in sect.~\ref{S:WDWscalar}. We
can reason as in sect.~\ref{SS:fastroll} that tunneling to an accelerated expansion
phase is favored since otherwise the barrier does not end. Since an accelerated
expansion phase requires that the potential is flat enough, it reasonable to expand
to leading order in an expansion in $V'/V$, and as argued previously it is also
reasonable to assume that the moduli are near a critical point in the potential.
Thus the arguments of sect.~\ref{S:thermal} that lead to eq.~(\ref{vmax}) can be
used. The new probability maximization conditions are
\begin{eqnarray}
\frac{1}{V}\frac{\partial V}{\partial\phi^{i}} & \simeq& 0
\label{extremiz1} \\
V(\phi;\beta)& \simeq & 5\nu(\phi),
 \label{extremiz2}
\end{eqnarray}
with
\begin{equation}
\nu=\frac{n_{dof}}{9\pi^{2}c^{4}}\frac{M_{s}^{4}}{M_{p}^{4}}=
\frac{n_{dof}}{9\pi^{2}c^{4}}(T)^{-3}(S_{R})^{-1}.
 \label{nu2}
\end{equation}
Equations (\ref{extremiz1}) (one for each modulus) determine, for any set of
parameters $\beta$, a region in moduli space for which the slow roll conditions are
satisfied. Of course, there may be some values of $\beta$ for which there is no
solution. Equation (\ref{extremiz2}) is then a further constraint that restricts
the parameter values only to those that satisfy it.

Equations (\ref{extremiz1})--(\ref{nu2}), however, leave the system
under-determined. The only exception is when there is only one parameter in the set
$\beta$. The logarithm of the WKB wave function at the point of emergence from the
barrier into the Lorentzian spacetime depends on the values of the moduli at this
point $\{S^L,T^L,z^L\}\equiv\{\phi_i^L\}$ and on $a_+$,
\begin{equation}
a_{+}=\frac{1}{2V(\phi)}\left[1+\sqrt{1-4\frac{\nu(S,T)}{V(\phi)}}\right].
 \label{aplus}
\end{equation}
Hence, it takes the following form in general
 \begin{equation}
\Phi=\Phi(S^L,T^L,z^L,a_{+},\beta ).
 \label{Phi}
\end{equation}
The wave function should be extremized with respect to all the moduli as well as
the parameters $\beta$. Since $\Phi$ depends on $\beta$ only through its dependence
on the potential, we have
\begin{eqnarray}
\frac{\partial\Phi}{\partial\beta}& = &\frac{\partial\Phi}{\left.\partial
V\right|}_{\phi_i^L} \frac{\partial V}{\left.\partial \beta\right|}_{\phi_i^L}=0,
\label{ext1} \\
\frac{\partial\Phi}{\partial\phi^i_L} & = & \frac{\partial\Phi}
{\left.\partial
a_{+}\right|}_{\phi_i^L} \frac{\partial a_{+}}{\partial\phi_i^L} +
\frac{\partial\Phi}{\left.\partial\phi^i_L\right|}_{a_{+}}=0.
 \label{ext2}
\end{eqnarray}

In principle, these equations determine the tunneling end-point in the moduli space
that serves as the initial values for the classical evolution of all the moduli.
They also determine the discrete parameters, the fluxes, gauge group parameters,
etc.. These equations therefore determine the particular flux configuration to
which the tunneling occurs. Of course, in general there may be more than one
solution to these conditions, so their solution may be a multiverse rather than a
universe.

In practice we used the particular expression for $\Phi$ that was obtained in
eq.~(\ref{elliptic}) for the case of a constant potential (CC) by arguing that the
tunneling took place to a flat point on the potential. Thus we argued that even
with the moduli, the equations of sect.~\ref{S:thermal} could be used with the
appropriate replacements for $\lambda$ and $\nu$. Equation (\ref{extremiz2}) then
follows from imposing $\partial\Phi/\partial V=0$ which implies that the whole set
of equations (\ref{ext1}), one for each parameter in the set $\beta$, are
satisfied. Clearly this is a sufficient but not a necessary condition for
extremization with respect to $\beta$. It would be interesting to explore the
nature of the general solutions to the extremization equations.

In the next section we will investigate the nature of the restrictions that we have
found for the specific  potentials for moduli that have been suggested recently by
KKLT.

\section{Application of the thermal boundary condition to the KKLT
model}

 \label{S:KKLT}

We would like to determine  more accurately the point on the moduli potential to
which the universe tunnels. For this we need to input some additional information
on the properties of the moduli potential. We will use here the KKLT model
\cite{Kachru:2003aw}. In this model, the potential has two contributions
$V_{SUGRA}$ and $V_{\overline{D}}$. The first takes the form of an $N=1$ SUGRA
potential (using the same units as before) with a Kaehler potential given by
eq.~(\ref{kaehler}) and a superpotential
\begin{equation}
W=A+BS+Ce^{-aT}.
 \label{superpot}
\end{equation}
The first two terms in eq.~(\ref{superpot}) come from the fluxes
\cite{Giddings:2001yu}, $A$ and $B$ are functions of the complex structure moduli,
and there is only one Kaehler modulus $T$. The third term in this expression can
arise from gaugino condensation in a gauge group living on a stack of 7-branes
wrapping a four cycle on the compact manifold. KKLT assume that it is possible to
ignore the third term and integrate out $S$ and the complex structure moduli,
assuming a flux configuration which makes their masses heavy. While this is not
strictly correct (see \cite{deAlwis:2005tf}) the corrections are not important for
the current discussion so we will ignore them. Then $S$ and the complex structure
moduli are constants, and the  effective superpotential is of the form
$W=W_{0}+Ce^{-aT}$ with a Kaehler potential $K=-3\ln(T+\bar{T})$ giving
\begin{equation}
V_{SUGRA}=\frac{aCe^{-aT_{R}}}{2T_{R}^{2}}\left(W_{0} +
(\frac{1}{3}T_{R}a+1)Ce^{-aT_{R}}\right).
\end{equation}
 The minimum of $V_{SUGRA}$ is at $DW=\partial_TW+\partial_TKW=0$, and
for consistency with the assumption that the volume is large, and therefore that
ten dimensional supergravity is valid, and for consistency with the expansion in
non-perturbative terms, one needs $T_{R}\gg1,\, aT>1$, so that we need to have
$W_{0}<1$. This can be achieved by fine tuning the fluxes.

The second contribution to the KKLT potential $V_{\overline{D}}$ comes from
anti-$D_{3}$ branes and breaks supersymmetry explicitly from the 4D perspective. It
is proportional to $\frac{1}{T_{R}^{3}}$ in a naive calculation of the anti-Dbrane
tension, however, a term proportional to $\frac{1}{T_{R}^{2}}$ has also been
proposed in the literature. The total potential has a shallow positive minimum at a
largish value of $T_{R}$, and a small barrier separates it from a positive tail
that goes to zero as $\sim1/T_{R}^{3}$.

Now let us check whether the end point of the tunneling can be on the asymptotic
tail using the arguments of section \ref{SS:fastroll}. The field $T_{R}$ is not
canonically normalized. The canonically normalized field is $x$ where
$T_{R}=e^{\sqrt{\frac{2}{3}}x}$. So the asymptotic dependence of the full potential
on $x$ is $V\sim e^{-3\sqrt{\frac{2}{3}}x}$. Recall the discussion in
sect.~\ref{S:WDWscalar} where we concluded that the (Euclidean) cosmological
scaling solution of a canonically normalized scalar field $\phi$ with an
exponential potential $V= A e^{\alpha\phi}$ gives a power law dependence for the
scale factor $a(t)\sim t^{p_{a}}$, with $p_{a}=1/\alpha^{2}$. If $p_{a}>1$,
$\alpha<1$, the expansion is accelerated. Recall also that for $|\alpha|>\sqrt{3}$
the prefactor $A$ needs to be negative for a scaling solution to exist. In the case
that the potential goes like $1/T_{R}^{3}$ then $\alpha=\sqrt{6}$  or in the case
that the potential goes like $1/T_{R}^{2}$ then $\alpha=\sqrt{\frac{8}{3}}$, both
significantly above $1$. The prefactor $d$ is positive. Hence we can conclude that
these potentials do no lead to accelerated expansion.

The argument of section \ref{SS:fastroll} indicated that for the tunneling barrier
to be finite, one needed an accelerating scale factor. Our conclusion is therefore
that the tunneling end point cannot be on the asymptotic tail of the potential in
the region $T_R\gg 1$ where $V_{\overline{D}}$ dominates. A possible loophole in
the argument is that perhaps it is possible to find a transient solution that
includes a brief period of accelerated expansion. However, also in this case the
potential energy has to dominate the energy balance. Then, ignoring numerical
factors of order unity and since $S_{R}$ is fixed at a number of $O(1)$,
eq.(\ref{extremiz2}) leads to the condition
\begin{equation}
\frac{5n_{dof}}{9\pi^{2}c^{4}} = d.
 \label{kkltcondA}
\end{equation}
As we discuss below, this relationship cannot be satisfied since $d$ has to be very
small, and the l.h.s. is not particularly small. Hence it is not possible to
satisfy the extremization conditions even in this case.

If the anti-Dbrane term is not included then the tail of the potential is steeper,
and if a potential of the form $\widetilde{d}/T_R^2$ is added then
eq.~(\ref{kkltcondA}) is replaced by $\frac{5n_{dof}}{9\pi^{2}c^{4}} =
\widetilde{d} T_R$ which is even harder satisfy. Our conclusion is therefore valid
for these additional cases.

Let us now check whether the tunneling end point can be near the shallow minimum,
or near the maximum of the barrier separating the minimum from the asymptotic
region. The dimensionless radiation $\nu$ is given by eq.~(\ref{nu2}) and the
dimensionless potential by
\begin{equation}
V=V_{SUGRA}+\frac{d}{T_{R}^{3}}.
\end{equation}
At the two extrema, the two terms are comparable. In the event that the tunneling
occurs to an extremum, which we have argued is a reasonable approximation, the wave
function is extremized at points where $\nu$ and $V$ are related as in
eq.~(\ref{extremiz2}). In our case $S_{R}$ fixed at a number of $O(1)$.

Equation (\ref{extremiz2}) gives
\begin{equation}
\frac{5n_{dof}}{9\pi^{2}c^{4}}=\frac{aAT_{R}e^{-aT_{R}}}{2}
\left(W_{0}+(\frac{1}{3}T_{R}a+1)Ce^{-aT_{R}}\right)+d,
 \label{kkltcond}
\end{equation}
where we have ignored a factor of $O(1)$. The factor
$\frac{5n_{dof}}{9\pi^{2}c^{4}}$ is  about $O(10^{-1})$, and is not expected to
vary much with different flux choices.  On the other hand, for models with one
condensate (i.e. one non-perturbative term as in the original KKLT example) and a
small CC, both extrema are at quite large values of $aT_{R}$. For large values of
$a T_R$ the r.h.s. of eq.(\ref{kkltcond}) is much smaller than 1/10. We conclude
that eq.(\ref{kkltcond}) cannot be satisfied at either extremum. For example, let
us consider the choice of parameters in KKLT, where $a=.1$, $A=1$, $T_R\sim
100-150$ and $d=1\times 10^{-9}$. Although the normalization of the potential in
KKLT is somewhat different than ours, it is clear that the r.h.s of
eq.~(\ref{kkltcond}) is of order $10^{-9}$ for the maximum, and much smaller for
the minimum where the CC was tuned to be very small.

Our conclusion is that the quantum cosmology arguments would prefer a modified KKLT
model, with additional exponential terms in the superpotential to allow extrema
that are closer to the central region of moduli space where eq.~(\ref{extremiz2})
has a chance to be obeyed. Such models can be  constructed by having several
non-perturbative terms in the superpotential (see, for example
\cite{Blanco-Pillado:2004ns} and \cite{Brustein:2005xx}). Clearly, it is easier to
satisfy eq.~(\ref{extremiz2}) at a maximum, a saddle point, or a metastable minimum
rather than  at a global minimum.

\section{Conclusions}

\label{S:conclusions}

We have seen that the probability distribution obtained from the wave function of
the universe can provide an interesting and restrictive dynamical selection
principle  on the landscape of string solutions without reference to the anthropic
principle.

We have proposed that the universe emerged from the string era in a thermally
excited state above the HH vacuum, and determined, within a mini-superspace model,
the probability of tunneling to different points on the landscape. We have
clarified the significance of including a radiation term for the HH wave function
of the universe and have shown that the radiation term switches the roles of the
Hartle-Hawking and Linde-Vilenkin wave functions. We have found that the potential
energy of the tunneling end point from which the universe emerges and begins its
classical evolution, is determined by the primordial temperature, and that  this
starting point can be followed by some interesting dynamics.

We have found that a more accurate treatment, even within the mini-superspace
approximation, requires the inclusion of the moduli fields, and we have included
them. By taking into account some generic properties of the moduli potential we
then argue that the tunneling to the tail of the moduli potentials is disfavored,
that the most likely emergence point is near an extremum, and that this extremum is
not likely to be in the outer region of moduli space where the compact volume is
very large and the string coupling very weak.  Combined, these considerations
select a class of values of the flux parameters etc., that characterize a universe,
or a multiverse, in the landscape.

We explicitly demonstrated the applicability of our arguments for the KKLT model of
moduli stabilization.  We have determined that for the KKLT model the tunneling to
the tail of the potential or to the vicinity of the barrier that separates the
minimum from the asymptotic region is disfavored. Our quantum cosmology arguments
favor tunneling to an extremum of the potential that is close to the central region
of moduli space as might be obtained from generalizations of the original KKLT
model.

Finally, we might consider the relevance of the counting program of Douglas and
collaborators \cite{douglas1,douglas2,douglas3} to our arguments. We have
calculated a quantity that is analogous to the square of the tunneling amplitude in
the calculation of a decay rate of an unstable particle. To obtain the total
tunneling rate one needs also the density of final states. Perhaps the counting
program can supply the latter, so that a complete calculation of the relative
probabilities of finding one or another universe, or a certain subset of the
multiverse, is obtained by taking the product of the two factors. The number
density of solutions does not seem to influence much the preferred value of the CC,
because the tunneling amplitude is sharply peaked as a function of the CC, while
the number density of solutions is uniform as a function of the CC. Perhaps it is
more relevant to the preferred values of other dynamical parameters.

\section{Acknowledgments}

We thank Henry Tye for a discussion on \cite{Sarangi:2005cs}. We also wish to thank
Raphael Bousso, Alex Buchel, Aharon Davidson, Frederik Denef, Eduardo Guendelman,
Shamit Kachru, Rob Myers, Gary Shiu and Rafael Sorkin for useful comments. This
research is supported in part by the United States Department of Energy under grant
DE-FG02-91-ER-40672 and the Perimeter Institute. R.B thanks PI for hospitality, SdA
thanks the department of Physics, Ben-Gurion University for hospitality.

\section*{Appendix}

Here we explain the relation of our work to the calculations of Sarangi and Tye
\cite{Sarangi:2005cs}.

Sarangi and Tye  (ST) have argued that the wave function needs to be modified due
to decoherence effects. They argue that the fluctuations of the metric and of other
light fields should be integrated out and traced over. Their calculation is rather
involved, but the final result can be justified and related to that in
sect.~\ref{S:thermal} by the following simple argument.

ST are essentially computing the thermal partition function in a
Friedman-Robertson-Walker (FRW) background at some undefined temperature. It is
well known that this calculation gives a contribution to the effective action (free
energy) which is just the energy density of radiation
i.e. a term proportional to $T^{4}$ %
\footnote{ST effectively work with a thermal circle whose radius is
called $T$.
This should in fact be identified with the inverse temperature - so
in our notation their $T$ should be replaced by our $T^{-1}$. %
}. It is also well known that in the FRW background the temperature scales as the
inverse of the scale factor $a^{-1}$. So the only question is: what is the
temperature?

Let us work with the action (\ref{action}). The effect of the unobserved
fluctuations can be represented by the entropy $\Sigma$ within a horizon volume. An
order of magnitude estimate of the upper bound on this is (a related though not
identical calculation is performed by ST)
$$
\Sigma\sim\int d^{3}n=\left(\frac{H^{-1}}{l_{s}}\right)^{3} \sim
\left(\frac{1}{\sqrt{\Lambda}l_{s}}\right)^{3},
$$
where we have assumed as in \cite{Sarangi:2005cs} that the infrared cutoff is the
horizon size and the ultraviolet one is the string scale $l_{s}$. Also we  have
used the Friedman equation to estimate $H\sim1/\sqrt{\Lambda}$. Comparing to the
entropy of a thermal state  $S\sim\sigma^{3}a^{3}T^{3}$,  we expect $S$ and
$\Sigma$ to agree to within a time independent factor $c^{3}$. The factor $c$ may
be numerically large but is not expected to be parametrically large. More
importantly it is time independent. Equating them
$\left(\frac{1}{\sqrt{\Lambda}l_{s}}\right)^{3}=c^{3}\sigma^{3}a^{3}T^{3}$ we find
that the effective temperature is $T=1/(ca\sqrt{\lambda}l_{s})$ where $\lambda$ is
the dimensionless CC introduced after eq.~(\ref{action2}). The energy density
associated with this radiation is then given by eq.~(\ref{density}). ST effectively
have $c=1$. However this is not consistent with the effective field theory
mini-superspace starting point, since as can be seen from eq.~(\ref{aplusminus}),
at $a=a_{\pm}\sim1/(\lambda^{\frac{1}{4}})$, this would give a radiation energy
density that is greater than string scale. So for consistency one should have
$c^{4}>n_{dof}$ as discussed in sect.~\ref{S:thermal}.

\bibliographystyle{apsrev}

\end{document}